\documentclass[sigconf]{acmart}

\AtBeginDocument{%
  }

\begin{document}

\title[UNIQUE: A Unified Retrieval and Ranking System]{UNIQUE: A Unified Retrieval and Ranking System for Large-Scale Feed Recommendation}

\author{Zhuang Liu}
\authornote{Both authors contributed equally to this research.}
\authornote{Work done during an internship at Baidu.}
\affiliation{%
  \institution{Beihang University}
  \city{Beijing}
  \country{China}}
\email{SY2412308@buaa.edu.cn}

\author{Yongkang Fu}
\authornotemark[1]
\affiliation{%
  \institution{Baidu Inc.}
  \city{Beijing}
  \country{China}}
\email{fuyongkang@baidu.com}

\author{Zuodong Yang}
\affiliation{%
  \institution{Baidu Inc.}
  \city{Beijing}
  \country{China}}
\email{yangzuodong@baidu.com}

\author{Guangxing Chen}
\affiliation{%
  \institution{Baidu Inc.}
  \city{Beijing}
  \country{China}}
\email{chenguangxing@baidu.com}

\author{Zonggang Wu}
\affiliation{%
  \institution{Baidu Inc.}
  \city{Beijing}
  \country{China}}
\email{wuzonggang@baidu.com}

\author{Yuqi Lu}
\affiliation{%
  \institution{Baidu Inc.}
  \city{Beijing}
  \country{China}}
\email{luyuqi@baidu.com}

\author{Shouke Qin}
\affiliation{%
  \institution{Baidu Inc.}
  \city{Beijing}
  \country{China}}
\email{qinshouke@baidu.com}

\author{Shantao Li}
\authornote{Corresponding author.}
\affiliation{%
  \institution{Baidu Inc.}
  \city{Beijing}
  \country{China}}
\email{lishantao@baidu.com}

\author{Maolin Wang}
\authornotemark[3]
\affiliation{%
  \institution{Hong Kong Institute of AI for Science, City University of Hong Kong}
  \city{Hong Kong}
  \country{China}}
\email{morinwang@foxmail.com}

\renewcommand{\shortauthors}{Liu et al.}

\copyrightyear{2026}
\acmYear{2026}
\setcopyright{cc}
\setcctype{by}
\acmConference[RecSys '26]{20th ACM Conference on Recommender Systems}{September 27-October 02, 2026}{Minneapolis, MN, USA}
\acmBooktitle{20th ACM Conference on Recommender Systems (RecSys '26), September 27-October 02, 2026, Minneapolis, MN, USA}
\acmDOI{10.1145/3773078.3831936}
\acmISBN{979-8-4007-2284-4/2026/09}

\begin{abstract}
Industrial mobile feed systems rely on a retrieval-ranking pipeline to serve large-scale, heterogeneous, and fast-changing content under strict latency constraints. However, existing pipelines still suffer from two critical issues: hierarchical quantization instability in candidate retrieval and information loss between separated retrieval and ranking stages. These issues hurt long-tail and cold-start recommendation and complicate efficient serving. To address them, we present UNIQUE, a unified retrieval and ranking recommendation framework with single-layer flat quantization. UNIQUE integrates generative code-based retrieval and target-aware ranking into one early-fusion architecture, enabling end-to-end training under a shared representation while preserving efficient candidate generation. A balanced quantization mechanism is further introduced to mitigate codebook imbalance and improve long-tail representation. Offline experiments evaluate UNIQUE from both retrieval and ranking perspectives, while codebook analysis shows more balanced resource allocation than hierarchical quantization. We deploy UNIQUE in the homepage feed, discovery-page, and short-video recommendation scenarios of Mobile Baidu, serving large-scale real-world traffic. Online A/B tests achieve a 0.96\% gain in total watch duration and a 1.08\% gain in total distribution volume, with notable improvements for new users and highly active users. Serving measurements show 89 ms P99 latency and 44.23\% online inference MFU. These results show that UNIQUE provides a stable, efficient, and production-ready framework for unified retrieval and ranking in industrial recommendation.

\end{abstract}



\maketitle
\section{Introduction}

Industrial recommender systems serve large-scale and highly dynamic content streams under strict latency constraints. In Mobile Baidu, Baidu's flagship mobile application, homepage feed recommendation, discovery-page recommendation, and short-video recommendation all rely on retrieving a compact candidate set from a large item corpus before downstream ranking. This retrieval-ranking cascade is widely adopted in industry because it balances full-corpus coverage and fine-grained scoring efficiency \cite{huang2013dssm,song2022ecm,zhou2018deepnetworkclickthroughrate}. However, these scenarios involve heterogeneous content such as articles, images, and videos, and require the system to capture both long-term user interests and fast-changing feedback signals.

Generative retrieval has recently emerged as a promising direction for large-scale recommendation. By mapping items into discrete semantic codes, generative models can avoid directly modeling an extremely large item vocabulary and can potentially improve generalization for long-tail or cold-start items \cite{rajput2023recommendersystemsgenerativeretrieval,hu2025idssemanticsgenerativeframework}. Recent studies further improve semantic identifier learning through learnable tokenization or unified generative modeling \cite{wang2025learnableitemtokenizationgenerative,zhang2025killingbirdsstoneunifying}. Nevertheless, deploying generative retrieval in industrial recommendation systems remains challenging.

The first challenge is quantization instability. Hierarchical semantic-ID methods such as RQ-VAE can propagate coarse-assignment errors to later levels. Under skewed industrial traffic, head items may overuse code capacity, leaving long-tail items underrepresented and weakening code fidelity.

The second challenge is the separation between retrieval and ranking. In cascaded systems, the two stages are trained with different objectives and exchange information only through candidate lists. Retrieval thus cannot fully benefit from ranking supervision, while ranking is constrained by upstream candidate quality.

To address these challenges, we propose UNIQUE, a unified retrieval and ranking framework with end-to-end quantization. UNIQUE connects feedback-aware semantic code learning with an early-fusion backbone, jointly optimizing code-based retrieval and target-aware ranking. Since February 2026, UNIQUE has been deployed in Mobile Baidu's recommendation system, serving homepage feed, discovery-page, and short-video scenarios with a unified retrieval-ranking workflow. Online A/B tests show measurable gains in total watch duration and distribution volume under real-time serving constraints.

The main contributions of this work are summarized as follows:
\begin{itemize}
  \item We propose UNIQUE, a unified retrieval and ranking framework that jointly performs code-based candidate generation and target-aware scoring for industrial recommendation.
  \item We introduce a feedback-aware single-layer flat quantization mechanism, reducing hierarchical quantization instability and improving long-tail representation.
  \item We deploy UNIQUE in Mobile Baidu's production recommendation system and demonstrate measurable online gains across large-scale feed recommendation scenarios under real-time serving constraints.
\end{itemize}

\section{Related Work}

\subsection{Discriminative Retrieval}

Discriminative retrieval learns dense user and item representations and selects candidates through similarity matching, such as inner product. Matrix factorization \cite{baltrunas2011contextmf} is a foundational method in this line. In sequential recommendation, GRU4Rec \cite{hidasi2016sessionbasedrecommendationsrecurrentneural} models user behavior sequences with RNNs, while BERT4Rec \cite{sun2019bert4rec} and SASRec \cite{kang2018selfattentivesequentialrecommendation} use Transformer architectures and self-attention mechanisms to capture user interests. The ranking stage usually focuses on CTR prediction \cite{karatzoglou2013ltr,wang2025universalframeworkcompressingembeddings}, with early studies emphasizing feature interaction modeling, such as DCN \cite{wang2021dcnv2}, DeepFM \cite{guo2017deepfmfactorizationmachinebasedneural}, AutoInt \cite{song2019autoint}, and GDCN \cite{wang2023deeperlighterinterpretablecross}. Later, DIN \cite{zhou2018deepnetworkclickthroughrate} introduces target-aware attention, DIEN \cite{zhou2018deepevolutionnetworkclickthrough} incorporates GRU for stronger sequential modeling, and SIM adopts search-based mechanisms to handle long behavior sequences. Large language models (LLMs) have also been used to enhance retrieval and ranking representations, including RLMRec for improving item embeddings, LLMRec \cite{wei2024llmreclargelanguagemodels} for augmenting user-item interactions, and BAHE \cite{geng2024bahe} for using different LLM layers to generate behavior representations and perform CTR prediction. Despite their fine-grained matching ability, discriminative methods have inherent limitations. They rely on independent ID representations, which lead to vocabulary explosion and sparsity in large-scale item corpora. Their retrieval and ranking modules are usually optimized separately, causing substantial information loss during cross-stage transfer.

\subsection{Generative Retrieval}

Generative retrieval reformulates recommendation as a sequence generation task and replaces traditional ID representations with discrete semantic codes. HSTU \cite{zhai2024actionsspeaklouderwords} shows that model capacity can improve with parameter scaling, but still trains retrieval and ranking independently. TIGER \cite{rajput2023recommendersystemsgenerativeretrieval} constructs hierarchical semantic identifiers with residual-quantized autoencoders, while EAGER \cite{liu2025etegrec}, LETTER \cite{wang2025learnableitemtokenizationgenerative}, and ETEGRec \cite{liu2025generativerecommenderendtoendlearnable} further improve the generative paradigms. Nevertheless, existing generative methods have prominent limitations. RQ-VAE hierarchical quantization brings error accumulation, generative and discriminative modules are isolated, training relies excessively on item priors and ignores behavioral posteriors, and long-tail items lack adequate codebook representation. These defects restrict industrial generative retrieval performance.

\subsection{Multi-stage Joint Optimization and Unified Retrieval Frameworks}

To mitigate the limitations of multi-stage cascades, related methods, like RankFlow \cite{qin2022rankflow}, CoRR \cite{huang2023corr}, transfer information through global losses or knowledge distillation, but still rely on separated model architectures. UniGRF \cite{zhang2025killingbirdsstoneunifying} unifies the two stages within a generative framework, but remains limited to a purely generative paradigm and does not incorporate the fine-grained modeling ability of discriminative retrieval. To address these issues, we propose UNIQUE, which systematically integrates a unified early-fusion encoder, single-layer flat quantization, and a target-code-aware discriminative network for scalable retrieval and ranking.
\section{Methodology}

\subsection{Problem Formulation}

Let $\mathcal{U}$ and $\mathcal{I}$ denote the user set and item set, respectively. For each user $u \in \mathcal{U}$, we observe a chronological behavior sequence
\begin{equation}
  \mathcal{S}_u = \{i_1, i_2, \ldots, i_T\}, \quad i_t \in \mathcal{I},
\end{equation}
where each interaction may contain heterogeneous side information such as item category, textual tokens, behavior type, exposure signal, click feedback, and dwell-time statistics. The goal of industrial recommendation is to retrieve a compact candidate set from the full item corpus and then estimate fine-grained user feedback for candidate ranking.

Instead of treating retrieval and ranking as two isolated stages, UNIQUE learns a unified model that couples code-based candidate generation with target-aware scoring. Each item $i$ is assigned a semantic code $c_i \in \{1,\ldots,M\}$ from a codebook $\mathcal{C}=\{\mathbf{e}_j\}_{j=1}^{M}$, where $M$ is the codebook size and $\mathbf{e}_j \in \mathbb{R}^{d}$ is a code embedding. Given user history $\mathcal{S}_u$ and a candidate item $i$, the model optimizes two coupled objectives: predicting the semantic code of the next interacted item for retrieval and estimating multiple feedback targets for ranking,
\begin{equation}
  p(c_i \mid \mathcal{S}_u), \qquad
  \hat{\mathbf{y}}_{u,i} = \{\hat{y}^{\mathrm{rel}}_{u,i}, \hat{y}^{\mathrm{ctr}}_{u,i}, \hat{y}^{\mathrm{dur}}_{u,i}, \hat{y}^{\mathrm{comp}}_{u,i}, \ldots\}.
\end{equation}
Here $\hat{y}^{\mathrm{rel}}_{u,i}$, $\hat{y}^{\mathrm{ctr}}_{u,i}$, $\hat{y}^{\mathrm{dur}}_{u,i}$, and $\hat{y}^{\mathrm{comp}}_{u,i}$ denote relevance, click-through rate, normalized watch-duration, and completion predictions, respectively. The ellipsis represents other scenario-specific industrial objectives. We use $\mathcal{K}$ to denote the enabled objective set for a given scenario. This formulation allows the retrieval signal and ranking supervision to be learned within the same architecture.

\subsection{Overview of UNIQUE}

UNIQUE is a unified retrieval and ranking framework built on early fusion. It contains three core components: an End-to-End Quantization Network (EQN) that maps items into a single-layer semantic code space for retrieval, a User Interest Generation Network (UIGN) that encodes user profiles and multi-granularity behavior sequences, and a Target-Code-Aware Discriminative Network (TDN) that injects candidate targets for multi-task ranking.

The key design principle is early fusion. Rather than generating candidates first and sending them to an independent ranking model, UNIQUE places user history tokens, candidate item targets, and candidate code targets into one Transformer-based computation graph. A target-attention split mechanism enables target tokens to attend to the user prefix while preventing unnecessary target-target interaction. As a result, the model can compute retrieval-oriented code logits and ranking-oriented item scores in a single forward pass, while allowing ranking supervision to shape the shared user representation.

\begin{figure*}[t]
  \centering
  \includegraphics[width=\textwidth]{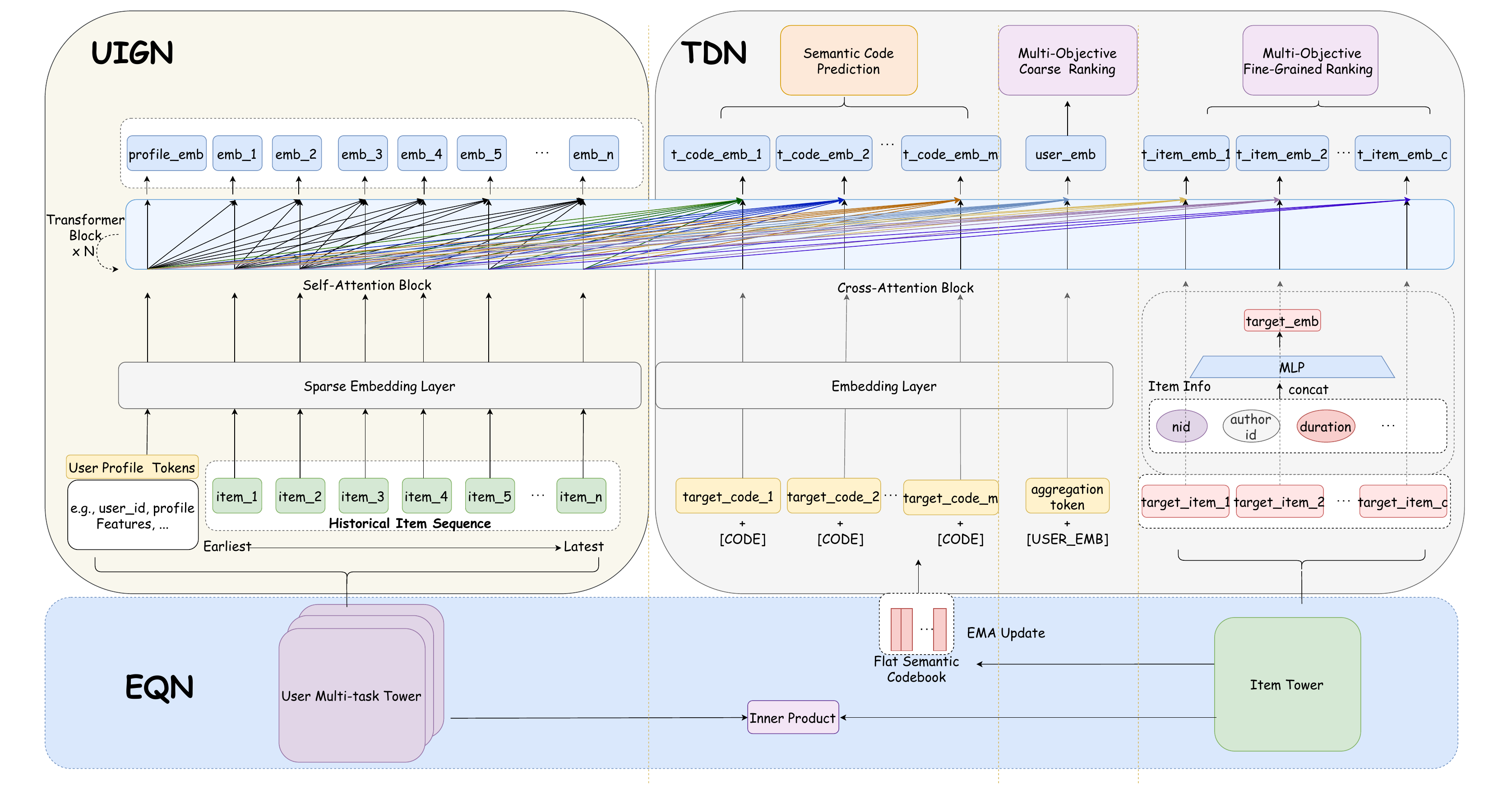}
  \caption{Overview of UNIQUE. The framework consists of three modules: UIGN encodes user profiles and multi-granularity behavior sequences; EQN learns feedback-aware flat semantic codes from item priors and behavioral posterior signals; and TDN injects candidate item/code targets into an early-fusion Transformer for code prediction and multi-objective ranking. Generation and discrimination losses jointly optimize retrieval and ranking under a shared representation.}
  \Description{UNIQUE contains UIGN for user interest modeling, EQN for feedback-aware semantic code learning, and TDN for target-code-aware prediction. User behavior, item targets, and code targets are fused in a Transformer to produce code retrieval logits and multi-objective ranking scores.}
  \label{fig:unique-architecture}
\end{figure*}

\subsection{End-to-End Quantization Network}

\subsubsection{Item Representation}

For each item $i$, EQN first constructs a dense item representation $\mathbf{z}_i \in \mathbb{R}^{d}$ from item-side features and feedback-aware signals. In practice, the item representation can combine prior semantic features, such as item ID, category, subcategory, publisher, title terms, and other metadata, with posterior behavioral signals encoded from user feedback. We denote this item-side encoding process as
\begin{equation}
  \mathbf{z}_i = f_{\theta}^{\mathrm{code\text{-}item}}(\mathbf{x}_i),
\end{equation}
where $\mathbf{x}_i$ represents the raw item feature fields and $f_{\theta}^{\mathrm{code\text{-}item}}$ is the item tower used for codebook learning. The resulting vector is normalized before quantization:
\begin{equation}
  \bar{\mathbf{z}}_i = \frac{\mathbf{z}_i}{\|\mathbf{z}_i\|_2}.
\end{equation}

\subsubsection{DSSM-Supervised Codebook Learning}

To make semantic codes reflect not only item-side priors but also posterior feedback from user interactions, EQN uses a DSSM-style multi-task two-tower network before quantization. The user side is encoded by task-specific user towers, and the item side is encoded by a shared code-update item tower:
\begin{equation}
  \mathbf{u}^{k}_{u}
  =
  f_{\theta}^{\mathrm{code\text{-}user},k}(\mathbf{x}_{u}),
  \qquad
  \mathbf{z}_{i}
  =
  f_{\theta}^{\mathrm{code\text{-}item}}(\mathbf{x}_{i}),
\end{equation}
where $k \in \mathcal{K}$ indexes an industrial feedback objective, such as relevance, click-through rate, watch duration, or short-video completion. The interaction score is computed by inner product:
\begin{equation}
  \label{eq:code-tower-score}
  s_{u,i}^{k}
  =
  \left\langle
  \mathbf{u}^{k}_{u},
  \mathbf{z}_{i}
  \right\rangle .
\end{equation}
During training, positive interacted items are paired with randomly rematched items from the batch to form contrastive supervision. Binary objectives such as relevance, CTR, and completion are optimized with binary feedback, while duration is optimized with regression-style feedback. The learned item representation $\mathbf{z}_{i}$ is then used as the input to vector quantization and EMA codebook update. In this way, the codebook is updated by item embeddings that have already absorbed user-feedback signals through the two-tower objectives, rather than by item metadata alone.

\subsubsection{Single-Layer Flat Quantization}

Unlike hierarchical quantization methods that assign multiple residual codes to each item, UNIQUE uses a single-layer flat codebook. The code index of item $i$ is obtained through nearest-neighbor search:
\begin{equation}
  c_i = \arg\min_{j \in \{1,\ldots,M\}}
  \|\bar{\mathbf{z}}_i - \mathbf{e}_j\|_2^2,
  \qquad
  \mathbf{q}_i = \mathbf{e}_{c_i}.
\end{equation}
This deterministic mapping removes the layer-wise approximation dependency in residual quantization. The quantized vector is passed to downstream modules using a straight-through estimator:
\begin{equation}
  \tilde{\mathbf{q}}_i =
  \bar{\mathbf{z}}_i + \mathrm{sg}(\mathbf{q}_i - \bar{\mathbf{z}}_i),
\end{equation}
where $\mathrm{sg}(\cdot)$ denotes the stop-gradient operation. This preserves gradients for the item encoder while using discrete semantic codes in the forward computation.

\subsubsection{Codebook Update and Balanced Usage}

The codebook is updated with an exponential moving average. Let $\mathcal{B}_j$ be the set of item representations in the current mini-batch assigned to code $j$. The code vector is updated by
\begin{equation}
  \mathbf{e}_j \leftarrow
  \gamma \mathbf{e}_j +
  (1-\gamma)
  \frac{1}{|\mathcal{B}_j|}
  \sum_{\bar{\mathbf{z}}_i \in \mathcal{B}_j} \bar{\mathbf{z}}_i,
\end{equation}
where $\gamma$ is the EMA decay factor. For unused codes, the previous code vector is retained. We also monitor codebook usage through the empirical distribution over assigned codes. Its perplexity is computed as
\begin{equation}
  \mathrm{PPL} =
  \exp\left(-\sum_{j=1}^{M} p_j \log(p_j+\epsilon)\right),
\end{equation}
where $p_j$ is the batch-level usage probability of code $j$.

To mitigate the Matthew effect in code assignment, UNIQUE further encourages balanced codebook usage. Let $r_j = p_j/\bar{p}$ be the usage frequency of code $j$ relative to the average $\bar{p}=1/M$, where $p_j$ is its batch-level usage probability, so $r_j>1$ marks an overused head code and $r_j<1$ an underused tail code. We adjust the assignment distance as
\begin{equation}
  d_j(\mathbf{z}) =
  \|\bar{\mathbf{z}}-\mathbf{e}_j\|_2^2 \cdot a_j,
  \qquad
  a_j = r_j^{\,\tau},
\end{equation}
where $\tau \ge 0$ sets the balancing strength ($\tau=0$ recovers standard nearest-neighbor assignment). Thus $a_j$ raises the assignment cost of overused head codes and lowers it for underused tail codes, discouraging popular items from collapsing into a few codes and improving the representation fidelity of long-tail items.

\subsection{User Interest Generation Network}

\subsubsection{Multi-Granularity Behavior Modeling}

UIGN models user interests from multiple behavior granularities. For each user, we construct a short-term sequence for recent interactions, a medium-term sequence for more stable preferences, and an interest sequence that summarizes semantically related historical behaviors. Each behavior token is represented by concatenating multiple feature embeddings, including item identifiers, categories, behavior types, temporal features, and feedback-related fields. These concatenated embeddings are projected into a shared hidden space:
\begin{equation}
  \mathbf{h}_t = f_{\theta}^{\mathrm{seq}}(\mathbf{x}_t), \quad
  \mathbf{h}_t \in \mathbb{R}^{d}.
\end{equation}
We denote the encoded short-term, medium-term, and interest behavior sequences as $\mathbf{H}^{\mathrm{st}}_u$, $\mathbf{H}^{\mathrm{mt}}_u$, and $\mathbf{H}^{\mathrm{int}}_u$, respectively, and denote the projected static user profile token as $\mathbf{h}^{\mathrm{user}}_u$. The final prefix sequence is
\begin{equation}
  \mathbf{X}_u =
  [\mathbf{h}^{\mathrm{user}}_u;
   \mathbf{H}^{\mathrm{st}}_u;
   \mathbf{H}^{\mathrm{mt}}_u;
   \mathbf{H}^{\mathrm{int}}_u]
  \in \mathbb{R}^{L \times d}.
\end{equation}

\subsubsection{Unified Early-Fusion Encoder}

The prefix sequence is encoded by a multi-layer Transformer encoder. To support joint generative and discriminative learning, UNIQUE appends target item embeddings and target code embeddings to the user prefix. We denote the resulting sequence as
\begin{equation}
  \mathbf{Z} =
  [\mathbf{X}_u; \mathbf{V}_i; \mathbf{R}_c],
\end{equation}
where $\mathbf{V}_i$ contains candidate item target tokens and $\mathbf{R}_c$ contains candidate code target tokens.
For each candidate item $i$, its target token $\mathbf{v}_i \in \mathbf{V}_i$ is obtained by applying an item-target embedding layer to the candidate item's raw feature fields, including item ID, category, content metadata, and other serving-time item features. The corresponding semantic code embedding can be appended through $\mathbf{R}_c$, so that candidate-level item features and code-level semantic information are fused in the same encoder.

The target attention split separates prefix self-attention from target-to-prefix attention. Prefix tokens perform self-attention over the user history. Target tokens attend to the prefix tokens to extract user-conditioned representations, while target-target attention is restricted. Formally, for a target token $\mathbf{t}$, the attention output is computed over the prefix:
\begin{equation}
  \mathrm{Attn}(\mathbf{t}, \mathbf{X}_u)
  =
  \mathrm{softmax}
  \left(
    \frac{\mathbf{W}_q\mathbf{t}
    (\mathbf{W}_k\mathbf{X}_u)^\top}{\sqrt{d}}
  \right)
  \mathbf{W}_v\mathbf{X}_u.
\end{equation}
This design allows all candidate targets to be evaluated in parallel while preserving the causal interpretation of generation from user history.

\subsubsection{Code Prediction}

For code prediction, semantic codes are treated in the same way as candidate items at the input layer. Each code $c_j$ is represented by a code target token $\mathbf{r}_j \in \mathbf{R}_c$ and is fed into the shared encoder together with the user prefix. The output at the corresponding code-token position is a user-code crossed representation:
\begin{equation}
  \mathbf{o}^{\mathrm{code}}_{u,j}
  =
  F_{\theta}^{\mathrm{enc}}(\mathbf{X}_u, \mathbf{r}_j),
  \qquad
  \ell_j
  =
  f_{\theta}^{\mathrm{code}}(\mathbf{o}^{\mathrm{code}}_{u,j}),
\end{equation}
where $f_{\theta}^{\mathrm{code}}(\cdot)$ is an MLP prediction head shared across code targets. The logits of all candidate codes are normalized to obtain the retrieval distribution:
\begin{equation}
  p(c=j \mid \mathcal{S}_u) =
  \frac{\exp(\ell_j)}
  {\sum_{m=1}^{M}\exp(\ell_m)}.
\end{equation}
The top-ranked codes under this distribution are selected for candidate generation. This design differs from conventional codebook-matching generative retrieval: the code logits are produced from early-fused user-code crossed vectors rather than from the similarity between a standalone user vector and code embeddings.

\subsection{Target-Code-Aware Discriminative Network}

TDN performs fine-grained discrimination over candidate targets within the same early-fusion architecture. For a candidate item $i$, its target representation is produced by the Transformer after attending to the user prefix:
\begin{equation}
  \mathbf{o}_{u,i} = F_{\theta}^{\mathrm{enc}}(\mathbf{X}_u, \mathbf{v}_i).
\end{equation}
This representation already contains user history, candidate item features, and code-aware semantic information. We feed it into task-specific prediction heads:
\begin{equation}
  \hat{y}^{k}_{u,i} =
  \sigma(f_{\theta}^{k}(\mathbf{o}_{u,i})),
  \quad
  k \in \mathcal{K}.
\end{equation}
The relevance head captures whether the user is likely to interact with the item, the CTR head estimates click probability under exposure, the duration head predicts normalized watch duration, and scenario-specific heads model signals such as short-video completion. These heads share the early-fused representation while keeping task-specific parameters, allowing fine-grained ranking supervision to improve the shared encoder.

\subsection{Training Objective}

UNIQUE is trained with both generative and discriminative objectives. For code generation, the target code $c_i$ is obtained from EQN, and the model minimizes the cross-entropy loss:
\begin{equation}
  \mathcal{L}_{\mathrm{gen}}
  =
  -\log p(c_i \mid \mathcal{S}_u).
\end{equation}

For discriminative learning, we use task-specific losses. Binary feedback tasks such as relevance, CTR, and completion are optimized with weighted binary cross-entropy, where $\mathcal{K}_{\mathrm{bin}} \subseteq \mathcal{K}$ denotes the enabled binary objectives:
\begin{equation}
  \mathcal{L}_{k}
  =
  -w_{u,i}^{k}
  \left[
    y_{u,i}^{k}\log \hat{y}_{u,i}^{k}
    +
    (1-y_{u,i}^{k})\log(1-\hat{y}_{u,i}^{k})
  \right],
  \quad k \in \mathcal{K}_{\mathrm{bin}}.
\end{equation}
Continuous feedback tasks such as watch duration are optimized with mean squared error:
\begin{equation}
  \mathcal{L}_{\mathrm{dur}}
  =
  w_{u,i}^{\mathrm{dur}}
  (y_{u,i}^{\mathrm{dur}}-\hat{y}_{u,i}^{\mathrm{dur}})^2.
\end{equation}
In industrial training, unsatisfactory clicks can be down-weighted to reduce noisy positive feedback. The overall discriminative loss sums over all enabled objectives:
\begin{equation}
  \mathcal{L}_{\mathrm{disc}}
  =
  \sum_{k \in \mathcal{K}}
  \lambda_k \mathcal{L}_k .
\end{equation}

For feedback-aware codebook learning, the DSSM-style code-update towers are trained with the same industrial feedback signals. Let $\mathcal{L}_{\mathrm{code}}^{k}$ denote the objective-specific loss computed from the two-tower scores in Eq.~\eqref{eq:code-tower-score}. The codebook-learning loss is
\begin{equation}
  \mathcal{L}_{\mathrm{code}}
  =
  \sum_{k \in \mathcal{K}}
  \mu_k \mathcal{L}_{\mathrm{code}}^k .
\end{equation}

The final objective combines the discriminative loss, code generation loss, feedback-aware codebook-learning loss, and quantization regularization:
\begin{equation}
  \mathcal{L}
  =
  \mathcal{L}_{\mathrm{disc}}
  +
  \alpha \mathcal{L}_{\mathrm{gen}}
  +
  \eta \mathcal{L}_{\mathrm{code}}
  +
  \beta \mathcal{L}_{\mathrm{quant}},
\end{equation}
where $\mathcal{L}_{\mathrm{quant}}$ denotes auxiliary quantization regularization, such as commitment loss or codebook-usage regularization, and $\alpha,\eta,\beta$ control the relative weights.

\subsection{Online Inference}

At inference time, UNIQUE first encodes the user prefix sequence and predicts a distribution over semantic codes. The top-ranked codes are mapped back to candidate items through the code-to-item index. Because multiple items may share the same semantic code, the mapping can retrieve multiple candidates per code according to business constraints such as freshness, popularity, or availability. The candidate items are then inserted as target tokens and evaluated by TDN to obtain relevance, CTR, duration, completion, and other scenario-specific scores. The final ranking score is computed by a configurable fusion function over these predictions, and the top-$K$ items are returned.

This inference procedure preserves the efficiency of code-based retrieval while retaining the fine-grained discrimination needed for industrial ranking. More importantly, both retrieval and ranking are served by the same early-fusion representation, improving serving efficiency and avoiding the information loss caused by conventional cascaded retrieval-ranking systems.

\section{Experiments}

We evaluate UNIQUE through offline experiments and an online A/B test. The evaluation is organized around three research questions:

\textbf{RQ1: Overall effectiveness.}
Can UNIQUE improve candidate generation while maintaining strong discriminative ranking accuracy?

\textbf{RQ2: Design contribution.}
How do the unified early-fusion encoder, single-layer flat quantization, target-attention split, and generative supervision contribute to the overall performance?

\textbf{RQ3: Industrial deployability.}
Can UNIQUE bring measurable online gains under real-time serving constraints?

\subsection{Experimental Setup}

\subsubsection{Datasets}

We evaluate UNIQUE on KuaiRand-Pure, a public video recommendation benchmark with rich feedback signals. Table~\ref{tab:dataset-statistics} summarizes the dataset statistics.

\begin{table}[!htbp]
  \caption{Statistics of experimental datasets.}
  \label{tab:dataset-statistics}
  \centering
  \begin{tabular}{lrrrr}
    \toprule
    Dataset & \#Users & \#Items & \#Inters. & Avg. Len. \\
    \midrule
    KuaiRand-Pure & 27,285 & 7,583 & 1,186,059 & 43.47 \\
    \bottomrule
  \end{tabular}
\end{table}

KuaiRand-Pure \cite{gao2022kuairand} provides video recommendation interactions with richer feedback signals and is used to evaluate both retrieval and ranking. We focus on this benchmark because its video-feed setting and multi-feedback interactions are closer to the industrial mobile feed scenarios studied in this work than rating-only datasets. Interactions are sorted chronologically for each user, and users with fewer than three interactions are filtered.

\subsubsection{Industrial Platform}

In addition to the public benchmark, UNIQUE is evaluated in the production recommendation system of Mobile Baidu, Baidu's flagship mobile application. Since January 2026, UNIQUE has been deployed in the retrieval stage and serves all online requests in three core recommendation scenarios: homepage feed recommendation, discovery-page recommendation, and short-video recommendation. The homepage feed provides a mixed stream of news, posts and videos for long-term interest matching; the discovery page adopts a two-column exploratory layout to help users discover new categories, niche content, and emerging trends; and the short-video scenario focuses on immersive full-screen consumption with high-frequency feedback.

Across these scenarios, UNIQUE shares the same model architecture, while scenario-specific input features and training data are used to adapt to different user behaviors and content distributions. This design improves consistency and maintainability compared with maintaining separate retrieval models for each scenario.

\subsubsection{Baselines}

We compare UNIQUE with three groups of baselines. Sequential retrieval baselines include SASRec \cite{kang2018selfattentivesequentialrecommendation} and TIGER \cite{rajput2023recommendersystemsgenerativeretrieval}. Ranking baselines include DCN \cite{wang2021dcnv2} and DIN \cite{zhou2018deepnetworkclickthroughrate}. Generative and unified baselines include HSTU \cite{zhai2024actionsspeaklouderwords} and UniGRF-HSTU \cite{zhang2025killingbirdsstoneunifying}. These methods cover sequence modeling, feature interaction, and unified retrieval-ranking.

\subsubsection{Evaluation Protocols}

We follow the leave-one-out protocol for offline evaluation. For each user, the last interaction is used for testing, the second-to-last interaction is used for validation, and all earlier interactions are used for training. To avoid candidate sampling bias in retrieval evaluation, we rank against the full item corpus and report Hit Ratio (HR@K) and NDCG@K.

For ranking evaluation, we report AUC on the click-through-rate prediction task. This setting ensures a fair comparison with ranking baselines, most of which are designed for CTR prediction and do not model relevance and watch duration simultaneously. The additional objectives in UNIQUE are used as auxiliary industrial supervision, while CTR-AUC serves as the common ranking metric.

\subsubsection{Implementation Details}

For the public benchmark, we train all models with the same train/validation/test split and use the validation set for hyperparameter selection. For UNIQUE on KuaiRand-Pure, the user and item embedding dimension is 256, the semantic codebook size is 1,024, and the code embedding dimension is 64. We use this smaller codebook because KuaiRand-Pure contains 7,583 items; the public benchmark is mainly used as a sanity check for retrieval-ranking effectiveness rather than as evidence for large-scale codebook allocation. The Transformer encoder uses 3 layers, 4 attention heads, hidden size 512, GELU activation, and no dropout. We optimize the model with Adam using a learning rate of $1\times10^{-3}$ with exponential decay and a batch size of 2048. The maximum user sequence length is 501, consisting of one static user token, 100 short-term behavior tokens, 200 medium-term behavior tokens, and 200 long-term interest tokens. The short-term sequence keeps the most recent 100 impressions and includes both clicked and exposed-but-unclicked samples, capturing immediate positive and negative feedback. The medium-term sequence keeps 200 recent satisfied-watch samples to represent more reliable preference signals. The long-term interest sequence keeps 200 historical behaviors by grouping user interests by item category and taking the most recent $n$ samples per category after truncation. Unless otherwise specified, all offline results use the same evaluation protocol. In production, UNIQUE uses a 16,384-code single-layer codebook for the industrial item corpus.

\subsection{RQ1: Overall Effectiveness}

RQ1 examines whether UNIQUE can improve candidate generation without sacrificing ranking quality. Table~\ref{tab:offline-performance} reports retrieval and CTR-AUC metrics on KuaiRand-Pure.

\begin{table}[!htbp]
  \caption{Offline results on KuaiRand-Pure.}
  \label{tab:offline-performance}
  \centering
  \small
  \setlength{\tabcolsep}{3.2pt}
  \begin{tabular}{lccccc}
    \toprule
    Model & HR@10 & HR@50 & NDCG@10 & NDCG@50 & CTR-AUC \\
    \midrule
    \multicolumn{6}{l}{\emph{Sequential retrieval baselines}} \\
    SASRec & 0.0304 & 0.0847 & 0.0189 & 0.0310 & -- \\
    Tiger & 0.0380 & 0.0902 & 0.0205 & 0.0378 & -- \\
    \midrule
    \multicolumn{6}{l}{\emph{Discriminative ranking baselines}} \\
    DIN & -- & -- & -- & -- & 0.6812 \\
    DCN & -- & -- & -- & -- & 0.7037 \\
    \midrule
    \multicolumn{6}{l}{\emph{Generative and unified baselines}} \\
    HSTU & 0.0453 & 0.1418 & 0.0273 & 0.0412 & 0.7150 \\
    UniGRF-HSTU & 0.0511 & 0.1453 & 0.0298 & 0.0430 & 0.7168 \\
    UNIQUE & \textbf{0.0575} & \textbf{0.1510} & \textbf{0.0310} & \textbf{0.0473} & \textbf{0.7252} \\
    \bottomrule
  \end{tabular}
\end{table}

UNIQUE achieves the best performance across all retrieval and ranking metrics. Compared with the strongest retrieval baseline, UniGRF-HSTU, UNIQUE improves HR@10 by 12.52\%, HR@50 by 3.92\%, NDCG@10 by 4.03\%, and NDCG@50 by 10.00\%. The larger gains on HR@10 and NDCG@50 indicate better top-ranked accuracy and broader candidate quality.

UNIQUE also obtains the highest CTR-AUC, improving over UniGRF-HSTU by 0.0084 absolute AUC and over HSTU by 0.0102. This suggests that the unified retrieval-ranking architecture improves candidate generation without sacrificing ranking quality, as code prediction and target scoring share useful supervision.

\subsection{RQ2: Design Contribution}

RQ2 isolates the contribution of UNIQUE's key designs. We compare the full model with four variants. The w/o early fusion variant replaces the shared early-fusion encoder with a separated architecture: user behaviors are encoded first, and candidate item/code representations are injected only through late fusion before the prediction heads. The other variants replace the single-layer flat codebook with a hierarchical RQ-VAE-style codebook, remove the target-attention split, or disable the generative code-prediction loss. All variants keep the remaining training losses and optimization settings unchanged. Table~\ref{tab:ablation-study} reports retrieval and ranking metrics.

\begin{table}[t]
  \caption{Ablation study on KuaiRand-Pure.}
  \label{tab:ablation-study}
  \centering
  \small
  \setlength{\tabcolsep}{3.5pt}
  \begin{tabular}{lcc}
    \toprule
    Variant & HR@50 & CTR-AUC \\
    \midrule
    UNIQUE & 0.1510 & 0.7252 \\
    w/o early fusion & 0.1398 (-7.42\%) & 0.7164 (-1.21\%) \\
    w/o flat quantization & 0.1421 (-5.89\%) & 0.7190 (-0.85\%) \\
    w/o target-attention split & 0.1456 (-3.58\%) & 0.7207 (-0.62\%) \\
    w/o generative objective & 0.1374 (-9.01\%) & 0.7115 (-1.89\%) \\
    \bottomrule
  \end{tabular}
\end{table}

The ablation results show that all major components contribute to the final performance, but their effects differ across retrieval and ranking. Removing the generative objective causes the largest HR@50 drop, indicating that code prediction is the main driver of candidate generation quality. In contrast, its CTR-AUC drop is relatively small, suggesting that item scoring is still mainly supported by the target-aware ranking path. Removing the unified encoder hurts both metrics, showing that early fusion between user behavior, candidate items, and candidate codes is important for sharing supervision across retrieval and ranking.

The flat quantization ablation also leads to clear degradation, especially on HR@50. This confirms that replacing the flat codebook with a hierarchical design weakens candidate generation, likely due to error propagation and less balanced code usage. Removing the target-attention split has a smaller but consistent impact, which suggests that restricting target-target interaction helps the model evaluate candidate targets more cleanly while preserving parallel scoring efficiency.

\begin{figure}[!htbp]
  \centering
  \includegraphics[width=\columnwidth]{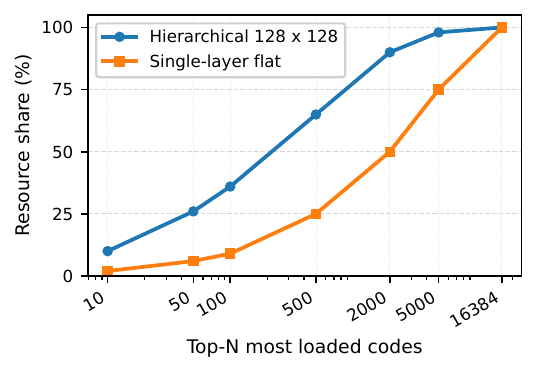}
  \caption{Cumulative resource share of the most heavily loaded codes under different codebook designs.}
  \Description{The figure compares hierarchical and single-layer flat codebooks. The hierarchical codebook concentrates a much larger share of resources in the top loaded codes, while the single-layer flat codebook distributes resources more evenly.}
  \label{fig:codebook-resource-share}
\end{figure}

We further analyze the production semantic codebook to understand why the flat quantization design matters at industrial scale. In the industrial implementation, the hierarchical $128 \times 128$ codebook is replaced by a single-layer codebook with the same total capacity of 16,384 codes. Since the available industrial analysis focuses on resource allocation, Figure~\ref{fig:codebook-resource-share} compares the cumulative share of resources assigned to the most heavily loaded codes.

The single-layer codebook substantially reduces resource concentration among head codes and lowers the resource-count variance from 1510.85 to 389.57, indicating a more balanced allocation of code capacity. Meanwhile, the average top-1 category share increases from 65.43\% to 78.08\%, suggesting that the resulting code clusters remain semantically coherent.

Together, these ablations test whether UNIQUE's gains come from the proposed unified architecture and codebook design rather than from a larger model or additional supervision alone.

\subsection{RQ3: Industrial Deployability}

RQ3 evaluates whether UNIQUE can translate offline improvements into online business gains under production-scale serving constraints. UNIQUE is deployed in the retrieval stage of Mobile Baidu's recommendation system and compared with the previous production retrieval pipeline through a one-week full online A/B test. The reported online gains pass the internal significance test with $p<0.05$. Table~\ref{tab:online-overall} reports the gains of core business metrics.

\begin{table}[t]
  \caption{Overall online A/B test results.}
  \label{tab:online-overall}
  \centering
  \begin{tabular}{lc}
    \toprule
    Metric & Relative Gain \\
    \midrule
    Total watch duration & +0.96\% \\
    Total distribution volume & +1.08\% \\
    User Retention Rate & +0.70\% \\
    \bottomrule
  \end{tabular}
\end{table}

The online results show that UNIQUE brings consistent gains on both engagement and traffic-side metrics. The improvement in total watch duration indicates that the retrieved and ranked items better match user interests, while the increase in distribution volume suggests that the unified model does not trade engagement quality for a smaller or more conservative exposure set. This is important for industrial deployment, where recommendation quality must improve without reducing traffic utilization.

Beyond the aggregate metrics, the A/B test also shows a significant improvement in user retention. UNIQUE accounts for 22.5\% of the retrieval candidate pool, indicating that the generated semantic-code candidates contribute meaningfully to the final serving results rather than acting as a marginal recall source.

We also break down the online gains by user activity level to examine whether UNIQUE improves different user groups consistently. High-, medium-, and low-activity users are defined by weekly active days of $>5$, 2--5, and $<2$, respectively; new users are those registered within 14 days. Table~\ref{tab:online-segments} reports the results.

\begin{table}[t]
  \caption{Online gains by user segment.}
  \label{tab:online-segments}
  \centering
  \begin{tabular}{lcc}
    \toprule
    Segment & Watch Time & Dist. Volume \\
    \midrule
    High-activity & +1.11\% & +1.29\% \\
    Medium-activity & +0.63\% & +0.65\% \\
    Low-activity & +0.29\% & +0.15\% \\
    New users & +1.44\% & +1.24\% \\
    \bottomrule
  \end{tabular}
\end{table}

The segmented results reveal two different sources of online improvement. First, UNIQUE achieves the largest watch-time gain for new users, suggesting that semantic code prediction helps alleviate cold-start sparsity when user histories are short. Second, highly active users also benefit substantially, indicating that the early-fusion encoder can exploit richer behavior sequences for fine-grained interest modeling. The gains are smaller for low-activity users, which is expected because both generative retrieval and discriminative ranking receive weaker historical signals in this group.

\begin{figure}[!t]
  \centering
  \includegraphics[width=0.85\columnwidth]{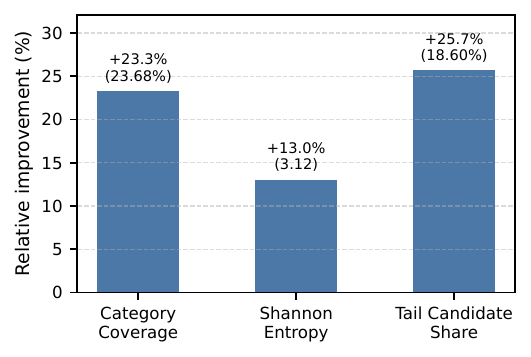}
  \caption{Relative improvements in online exposure diversity and long-tail coverage.}
  \Description{The bar chart shows the relative improvement of UNIQUE over the production baseline in category coverage, Shannon entropy, and tail candidate share.}
  \label{fig:diversity-relative-improvement}
\end{figure}

We further examine business-side diversity from online exposure logs. Figure~\ref{fig:diversity-relative-improvement} reports relative improvements over the production baseline in category coverage, per-user Shannon entropy, and tail candidate share among top-1K served candidates. Tail items are defined as the bottom 20\% of distributable items ranked by historical clicks, covering 1.82 million items in our online resource pool.

These online diversity results complement the codebook resource analysis. Higher category coverage indicates that exposed items span more second-level categories, while higher Shannon entropy suggests that each user's served candidate list is less concentrated in a few categories. UNIQUE reaches a 1.04\% exposure share for long-tail items, suggesting that it does not improve engagement by over-concentrating exposure on head items. Together, these results indicate that the flat and balanced codebook helps preserve diverse semantic regions in production candidate generation.

Finally, we evaluate serving efficiency under real-time production constraints. At peak traffic, the retrieval service handles approximately 10,000 QPS with an average latency of 37 ms. The inference cost is approximately \$0.0013 per 1,000 requests, supporting cost-effective large-scale deployment. Table~\ref{tab:system-efficiency} reports latency, MFU, and cache effectiveness. During peak traffic, the online system uses a 15-second cache window at noon and a 30-second cache window in the evening.

\begin{table}[t]
  \caption{Online system efficiency.}
  \label{tab:system-efficiency}
  \centering
  \begin{tabular}{lc}
    \toprule
    Metric & Value \\
    \midrule
    P80 latency & 35 ms \\
    P99 latency & 89 ms \\
    Online inference MFU & 44.23\% \\
    Noon cache hit rate (15s window) & 37\% \\
    Evening cache hit rate (30s window) & 60\% \\
    \bottomrule
  \end{tabular}
\end{table}

The recommendation service runs on approximately 400 NVIDIA L20 GPUs, with each service instance bound to a single GPU. We employ TensorRT mixed-precision (FP16) inference to lower GPU memory footprint and boost inference throughput. During the retrieval stage, UNIQUE generates $N=100$ semantic codes to retrieve all related items. The retrieved items (totaling hundreds of thousands) are ranked by the inner product between user and item embeddings in the coarse ranking stage, and the top 10k are grouped by code then fed into the fine-grained ranking stage to produce $M=30$ items per code for subsequent processing steps in the recommendation system.

These retrieval parameters reflect a quality-efficiency trade-off: the beam width $N$ controls code-level exploration (more semantic regions and better long-tail coverage at near-linear scoring cost), while the per-code size $M$ controls within-code exploitation. Online sensitivity analysis shows $N=100$, $M=30$ achieves the best balance: raising $N$ to 200 yields marginal diversity gains at roughly $2\times$ cost, while cutting $M$ to 10 significantly hurts recall. A more balanced codebook further prevents head codes from dominating the search space, so a moderate beam width suffices.

In our unified framework, a GPU inference engine integrates indexing and prediction to produce multi-objective user representations in real time. Item embeddings are precomputed hourly and synchronized through a message queue, and for cold-start freshness new items enter the training stream within 5 minutes of first impression and are quantized within 1 hour. A parameter server keeps UIGN and TDN versions consistent with negligible delay.

The system metrics show that UNIQUE remains compatible with real-time serving constraints: the P99 latency stays below 100 ms and the caching mechanism absorbs repeated high-traffic requests at peak hours. Overall, these results validate UNIQUE in real-world large-scale recommendation scenarios.

\section{Conclusion}

We presented UNIQUE, a unified retrieval and ranking framework for industrial mobile feed systems. By combining feedback-aware flat quantization with an early-fusion backbone, UNIQUE enables code-based candidate generation and target-aware scoring to share representations. Offline experiments, production codebook analysis, and online A/B tests on Mobile Baidu show consistent gains in retrieval quality, ranking accuracy, codebook balance, and business metrics, demonstrating the practicality of unified retrieval-ranking modeling under real-time serving constraints.

\begin{acks}
We thank our collaborators on Baidu's Feed Recall team, as well as Jianjia Zheng, Hongyang Wei, and Mengmeng Ge, for their invaluable support, constructive feedback, and insightful discussions throughout the development of this work.

During the preparation of this work, the authors used generative AI tools solely for language polishing and readability improvement. After using these tools, the authors reviewed and edited the content as needed and take full responsibility for the content of the publication.
\end{acks}

\bibliographystyle{ACM-Reference-Format}
\bibliography{refs}

\end{document}